\documentclass[aps,pra,preprint]{revtex4}

\usepackage{epsfig}
\usepackage{subfigure}

\begin{document}

\title{Mean field master equation for self-interacting baths II: extension to time-dependent subsystem Hamiltonians}
\author{Joshua Wilkie and Murat \c{C}etinba\c{s}}
\affiliation{Department of Chemistry, Simon Fraser University, Burnaby, British Columbia V5A 1S6, Canada}

\date{\today}

\begin{abstract}

A mean field argument is used to derive a master equation for systems simultaneously interacting with external fields and coupled environmental degrees of freedom. We prove that this master equation preserves positivity of the reduced density matrix. Solutions of the master equation are compared with exact solutions for a system consisting of three spins which is manipulated with a sequence of laser pulses while interacting with a spin-bath. Exact solutions appear to converge to the master equation result as the number of bath spins increases.

\end{abstract}

\maketitle

\section{Introduction}

Decoherence constitutes a potentially serious problem for a wide range 
of proposed technologies which exploit quantum phase interference. Examples
include molecular electronics, laser control of chemical reactions, quantum computing and molecular motors. Since condensed phase implementation of such schemes is probably necessary for nontrivial applications, some understanding of the influence of bath dynamics on system-bath interaction could prove useful for avoiding decoherence. Representation of condensed phase environments with the uncoupled oscillator baths commonly assumed as a starting point in older theories\cite{OLD} is not generally valid. Indeed, recent experimental\cite{Gasp} and numerical\cite{Miyano} evidence supports the notion that atoms and molecules in condensed phases exhibit chaotic dynamics and hence strong bath self-interaction. In addition, some numerical simulations suggest that baths which have strong self-interaction cause much less decoherence\cite{TW} than 
would be predicted by uncoupled oscillator models. Hence a theory of subsystem dynamics which can account for the effects of self-interacting baths could have important application in this area. Since lasers are often employed to initialize and 
manipulate such technologies the theory should also allow for interactions of the subsystem with external fields. 

We recently used a non-perturbative mean field approximation to derive
a non-Markovian positivity-preserving master equation for systems interacting with coupled baths\cite{JJW,JW,Wilk1,Wilk2}. Tests of this equation against exact results for a spin interacting with a coupled spin-bath 
showed good agreement\cite{JJW}. Here we extend the theory by allowing the subsystem to interact with 
external time-varying electromagnetic fields. 

A natural starting point for any theory of subsystem dynamics is the exact projection operator approach introduced by Nakajima and Zwanzig\cite{Zwan}. In the
case of time-independent subsystem Hamiltonians this leads to a unique (up to the definition of the projection operator) and exact integro-differential master equation. Unfortunately, the memory kernel which weights the contributions of the previous states of the subsystem to its future cannot be calculated in practice. Approximation of this kernel for a particular choice of projection operator formed the basis of our previous work\cite{JJW,JW,Wilk1,Wilk2}. The extension to time-dependent subsystem Hamiltonians requires a similar derivation from first principles. This is outlined in section II. The essential step, as in our
previous work\cite{JJW}, is an appropriate mean field approximation for the 
interactions of the subsystem and environment.

The reduced density matrix of the subsystem should be positive semi-definite but violation of this property is a common failing of many master equations\cite{Opp}. We therefore prove in section III that the master equation obtained in section II preserves positivity.

In section IV we introduce a model system which represents a set of three
qubits of a quantum computer which are manipulated with a sequence of laser
pulses. This three spin subsystem is allowed to interact with a bath of
strongly coupled spins designed to model solid state environmental modes. Solutions
are obtained for baths with varying numbers of spins. In section V we
discuss the method used to obtain solutions of the master equation. The
exact and master equation solutions are compared in section VI. We show that
the exact results approach the mean field results as the number of bath
modes increases.

\section{Mean field master equation}

Define a Nakajima--Zwanzig\cite{Zwan} projection operator $P$ on the total (system plus bath) density $\chi(t)$ such that 
\begin{equation}
P\chi(t)=\rho(t){\cal B},
\label{Pr}
\end{equation}
where $\rho(t)$ is the system density and ${\cal B}$ is the canonical bath density. Similarly, define $Q=1-P$. 

Consider a Hamiltonian
\begin{equation}
H(t)=H_s+{\cal E}(t)+H_b+\sum_{\mu}S_{\mu}R_{\mu}\label{HAM}
\end{equation}
where ${\cal E}(t)$ involves only system operators and represents the effects
of external electromagnetic fields. The operators $H_s$ and $S_{\mu}$ represent the system while $H_b$ and $R_{\mu}$ represent the bath. Define a time-independent operator
\begin{equation}
L=(1/\hbar)[H_s+H_b+\sum_{\mu}S_{\mu}R_{\mu},\cdot ~]
\end{equation}
which is the Liouville operator in the absence of external fields, and a time-dependent operator 
\begin{equation}
{\cal F}(t)=(1/\hbar)[{\cal E}(t),\cdot ~]
\end{equation}
for the lasers. Then using the facts that $P$ and ${\cal F}(t)$ commute and that $PQ=0$ it can be shown that
\begin{eqnarray}
d P\chi(t)/dt&=&-(i/\hbar) [PLP+{\cal F}(t)] ~P\chi(t) - (i/\hbar)PLQ~Q\chi(t)
\label{eq1}\\
d Q\chi(t)/dt&=&-(i/\hbar) [QLQ+{\cal F}(t)]~Q\chi(t) - (i/\hbar)QLP~P\chi(t)\label{eq2}.
\end{eqnarray}
Equation (\ref{eq2}) should now be solved for $Q\chi(t)$ and substituted into Eq. (\ref{eq1}). 

We have shown elsewhere\cite{JJW,Wilk2} that $P$ is non-Hermitian and that therefore $QLQ$ is non-Hermitian with a complex spectrum. Let $\omega_j$ and $\gamma_j$ denote the real and imaginary parts of an eigenvalue of $QLQ$ 
and let $|\phi_j)$ and $(\Phi_j|$ be the associated right and left eigenvectors\cite{WB}. We may expand $Q\chi(t)=\sum_j C_j(t)|\phi_j)$ in the complete eigenbasis and using orthonormality (i.e $(\Phi_j|\phi_k)=\delta_{j,k}$) it
follows that (\ref{eq2}) can be rewritten in the form
\begin{equation}
dC_j(t)/dt=(-i\omega_j-\gamma_j)C_j(t) -(i/\hbar)\sum_k (\Phi_j|{\cal F}(t)|\phi_k)C_k(t)- (i/\hbar)(\Phi_j|QLP~P\chi(t).\label{eq3}
\end{equation}
Now, the matrix elements $(\Phi_j|{\cal F}(t)|\phi_k)$ should be dominated by the overlap of the bath part of the generalized eigenstates. For a sufficiently large bath the overlap represents an integration over a product of two essentially random functions. On this basis we should expect $(\Phi_j|{\cal F}(t)|\phi_k)$ to vanish for $j\neq k$. Secondly, note that since ${\cal E}(t)$ is Hermitian, ${\cal F}(t)$ has a real spectrum which is symmetric about zero, and hence the diagonal elements $(\Phi_j|{\cal F}(t)|\phi_j)$ are zero on average. The first approximation in our derivation is thus to neglect the matrix elements of ${\cal F}(t)$ in Eq. (\ref{eq3}). Assuming an initial state of the form $\chi(0)=\rho(0){\cal B}$ we then obtain
\begin{equation}
C_j(t)=-(i/\hbar)\int_0^tdt' e^{(-i\omega_j-\gamma_j)(t-t')}(\Phi_j|QLP~P\chi(t')
\end{equation}
which in turn then gives the desired solution
\begin{equation}
Q\chi(t)=-(i/\hbar)\int_0^tdt' \sum_j e^{(-i\omega_j-\gamma_j)(t-t')}|\phi_j)(\Phi_j|QLP~P\chi(t')\label{eq4}
\end{equation}
which can be substituted into equation (\ref{eq1}).

Before making this substitution we require further approximation to eliminate the explicit dependence on the unknown generalized eigenvalues and eigenvectors. Our previous approximation exploited the large number of bath degrees of freedom and the consequent randomness of the generalized eigenvectors. The large number of bath modes also implies a large spectral density of states for $\omega_j$ and $\gamma_j$. A large number of terms will thus contribute to the sum in
(\ref{eq4}) suggesting that perhaps the sum can be replaced by its average.
That is, suppose that
\begin{eqnarray}
\sum_j e^{(-i\omega_j-\gamma_j)t}|\phi_j)(\Phi_j| \simeq \langle e^{(-i\omega-\gamma)t}\rangle \langle \sum_j |\phi_j)(\Phi_j| \rangle=W(t) {\bf 1}
\end{eqnarray}
where $W(t)=\langle \cos\omega t e^{-\gamma t} \rangle$ and the angle brackets denote an average over the generalized spectral density. Here we have used the fact\cite{JJW,Wilk2} that the spectral density is symmetric under $\omega\rightarrow -\omega$ and the closure relation for the generalized eigenvectors. This mean field type approximation, which should be accurate for sufficiently large baths, implies that
\begin{equation}
Q\chi(t)=-(i/\hbar)\int_0^tdt' ~W(t-t')QLP~P\chi(t').\label{eq5}
\end{equation}

Substituting (\ref{eq5}) into Eq. (\ref{eq1}) and tracing over bath degrees of freedom it can then be shown\cite{JJW} that
\begin{eqnarray}
d\rho(t)/dt&=&-(i/\hbar)[H_s+{\cal E}(t)+\sum_{\mu}\bar{R}_{\mu}S_{\mu},\rho(t)]\nonumber \\
&-&(1/\hbar^2)\sum_{\mu,\nu}C_{\mu,\nu}\int_0^tdt' ~W(t-t')\{[\rho(t')S_{\nu},S_{\mu}]+
[S_{\nu},S_{\mu}\rho(t')]\},
\label{masterb}
\end{eqnarray}
where $\bar{R}_{\mu}={\rm Tr}_b\{R_{\mu}{\cal B}\}$ and $C_{\mu,\nu}={\rm
  Tr}_b\{(R_{\nu}-\bar{R}_{\nu})(R_{\mu}-\bar{R}_{\mu}){\cal B}\}$
denote canonical (i.e. ${\cal B}=e^{-H_b/kT}/{\rm Tr}_b\{e^{-H_b/kT}\}$) averages and variances of bath operators. This is essentially the same master equation derived in Ref. \cite{JJW} except the subsystem Hamiltonian is now time dependent. Note that $W(t)$ plays the role of a memory function: it weights the contributions of previous states of the system.

A careful treatment of the spectral properties of ${\cal A}=QLQ$ allows one to calculate the mean spectral density which in turn can be used to calculate $W(t)$. Using this approach we have shown\cite{Wilk2} that
\begin{eqnarray}
W(t)&=&[1-\frac{4}{3\pi} (pt)^1+\frac{1}{8}(pt)^2-\frac{4}{45\pi}(pt)^3+\frac{1}{48}(pt)^4 ]e^{-(q t)^2/8}
\label{W2}
\end{eqnarray}
where
\begin{eqnarray}
p&=&[\langle {\cal A}{\cal A}^{\dag}\rangle-\langle {\cal
  A}{\cal A}\rangle]/\sqrt{\langle {\cal A}{\cal A}^{\dag}\rangle} \\
q&=&[\langle {\cal A}{\cal A}^{\dag}\rangle+\langle {\cal A}{\cal
  A}\rangle]
/\sqrt{\langle {\cal A}{\cal A}^{\dag}\rangle}.
\end{eqnarray}
The angle brackets here denote an average over the Liouville-Hilbert space i.e., for any operator $F$,
\begin{equation}
\langle F \rangle=\lim_{m,n\rightarrow \infty}(1/mn)\sum_{i=1}^m\sum_{j=1}^n (i,j|F|i,j) 
\end{equation}
where $|i,j)$ states denote a complete set.
Simplified formulas for the real parameters $\langle {\cal A}{\cal A}^{\dag}\rangle$  and $\langle {\cal
  A}{\cal A}\rangle$ are provided in Appendix A of Ref. \cite{JJW}. The memory function (\ref{W2}) is always positive and usually has a shape which is nearly gaussian.

Note that the assumptions we have made in the above derivation mean that the master equation (\ref{masterb}) is valid in the limit of large bath. 

\section{Proof of positivity}

Define operators $L(t)= (1/\hbar)[H_s+{\cal E}(t)+\sum_{\mu}\bar{R}_{\mu}S_{\mu},\cdot ~]$ and $L_d$ such that $L_d\rho=\sum_{\mu,\nu}C_{\mu,\nu}\{[\rho S_{\nu},S_{\mu}]+
[S_{\nu},S_{\mu}\rho ]\}$ and a function $M(t)=\tau \delta(t)-W(t)$ where $\tau=\int_0^{\infty}dt~W(t)$. We may then write (\ref{masterb}) in the form
\begin{equation}
d\rho(t)/dt=-\{iL(t)+\tau L_d\}\rho(t)+\int_0^tdt'M(t-t')L_d\rho(t').\label{raw}
\end{equation}
Rather than attempt to prove positivity for (\ref{raw}) we consider instead the related
equation
\begin{equation}
\partial \tilde{\rho}(t,s)/\partial t=-\{i(L(s)-i\partial/\partial s)+\tau L_d\}\tilde{\rho}(t,s)+\int_0^tdt'M(t-t')L_d\tilde{\rho}(t',s)\label{raw2}
\end{equation}
in which we have introduced a new variable $s$ which eliminates the explicit time dependence 
of $L$. [Note the similarity of this transformation to that employed in the $(t,t')$ method\cite{Pesk}.] We will now show that (\ref{raw2}) preserves positivity of $\tilde{\rho}(t,s)$, and
since $\rho(t)=\tilde{\rho}(t,s)|_{s=t}$, that (\ref{raw}) preserves positivity of $\rho(t)$.

Consider that both $L(s)$ and $i\partial/\partial s$ are Hermitian operators and that $L_d$ is of completely-positive-dynamical-semigroup\cite{dsg} form. It follows that the operator
${\cal D}=-i(L(s)-i\partial/\partial s)-\tau L_d$ is the generator of a completely-positive-dynamical-semigroup\cite{dsg}. Laplace transforming (\ref{raw2}) shows that $\tilde{\rho}(t,s)=T(t,s)\rho(0)$ where the propagator $T(t,s)$ is obtained by inverting the equation
\begin{eqnarray}
R(z,{\cal D}+\tilde{M}(z)L_d)&=&(z-{\cal D}-\tilde{M}(z)L_d)^{-1}\\
&=&\int_0^{\infty}dt~ e^{-zt}T(t,s) \label{resolve}
\end{eqnarray}
for the resolvent operator $R(z,{\cal D}+\tilde{M}(z)L_d)$. Here $\tilde{M}(z)$ is the Laplace transform of $M(t)$. 

To invert Eq. (\ref{resolve}) it is convenient to first show that $T(t,s)$ can be expressed in dynamical semigroup form. This means that there exists an operator ${\cal A}$ such that $T(t,s)=e^{{\cal A}t}$. Equivalently, this means that
the propagator can be written in terms of the resolvent via 
\begin{equation}
T(t,s)=\lim_{n\rightarrow\infty}\left[ \frac{n}{t}R(n/t,{\cal A})\right]^n
\end{equation}
which implies that $T(t,s)$ will preserve positivity if $R(z,{\cal A})=(z-{\cal A})^{-1}$ is positive for large real $z$. Since the resolvent $R(z,{\cal A})$ must be equivalent to the resolvent $R(z,{\cal D}+\tilde{M}(z)L_d)$ (i.e. the solution $\tilde{\rho}(t,s)$ is unique) it follows that $T(t,s)$ will preserve positivity if $R(z,{\cal D}+\tilde{M}(z)L_d)$ is positive for large real $z$. This can be readily proved as we show below. Thus, the essential step is showing that the solutions of (\ref{raw2}) can be written in dynamical semigroup form. Substituting
${\cal A}={\cal D}+\int_0^{\infty} dt'M(t')L_d\delta_{-t'}$ into $d\tilde{\rho}(t,s)/dt={\cal A}\tilde{\rho}(t,s)$ and using the boundary condition $\rho(t)=0$ for $t< 0$ one readily recovers Eq. (\ref{raw2}). [Here $\delta_{-t'}f(t)=f(t-t')$ is the delay operator. A detailed derivation of ${\cal A}$ is presented in Ref. \cite{Wilk2}.] Hence, the solutions of Eq. (\ref{raw2}) can indeed be written in dynamical semigroup form and we need only show that our original resolvent is a positive operator.

Clearly we may write
\begin{eqnarray}
R(z,{\cal D}+\tilde{M}(z)L_d)&=&R(z,{\cal D})\left(1-\tilde{M}(z)L_dR(z,{\cal D})\right)^{-1}\\
&=&R(z,{\cal D})\sum_{k=0}^{\infty}\left[ \tilde{M}(z)L_dR(z,{\cal D})\right]^k,
\end{eqnarray}
and since both $L_d$ and $R(z,{\cal D})=(z-{\cal D})^{-1}$ preserve positivity it follows that $T(t,s)$ will preserve positivity if $\tilde{M}(z)$ is positive for large real $z$.
Now, since 
\begin{eqnarray}
\tilde{M}(z)&=&\int_0^{\infty}dt ~e^{-zt}M(t)\\
&=&\tau-\int_0^{\infty}dt~ e^{-zt}W(t)\\
&=&\int_0^{\infty}dt ~W(t)- \int_0^{\infty}dt ~e^{-zt}W(t)\\
&=&\int_0^{\infty}dt~(1-e^{-zt})W(t)
\end{eqnarray}
it follows that $\tilde{M}(z)$ is positive if $W(t)$ is positive. In fact this is true for 
the memory function of equation (\ref{W2}). Hence, Eq. (\ref{raw2}) preserves positivity of 
$\tilde{\rho}(t,s)$.

Finally, define $\rho(t)=\tilde{\rho}(t,s)|_{s=t}$ from which it follows that
\begin{equation}
d\rho(t)/dt=\partial \tilde{\rho}(t,s)/\partial t|_{s=t}+\partial \tilde{\rho}(t,s)/\partial s|_{s=t}
\end{equation}
and inserting (\ref{raw2}) then gives our original equation (\ref{raw}). Thus, positivity
of $\rho(t)$ is preserved by the master equation (\ref{masterb}).

\section{Spin--Spin-Bath Model}

Our model system represents three qubits (i.e. two-level systems) which are manipulated with a sequence of laser pulses while the whole system interacts with an environment. Two of the qubits represent pairs of electronic states of impurities in a crystalline solid at low temperature. The third qubit represents two vibrational levels of an optical phonon mode which is
used as a means of transferring information from one impurity to the other via vibronic coupling induced by a laser. Other vibrational modes of the crystal (i.e. the environment) are represented by a number $n_s$ of coupled spin-1/2 modes. Obviously the representation of phonon
modes by spin-1/2 modes is valid only at low temperature. Here we set $kT=.0067$ eV which corresponds to liquid nitrogen temperature. 

\begin{figure}[htp]
\caption{Without dissipation}
\subfigure[$\rho_{00}^{(0)}$ and $\rho_{11}^{(0)}$]{
\epsfig{file=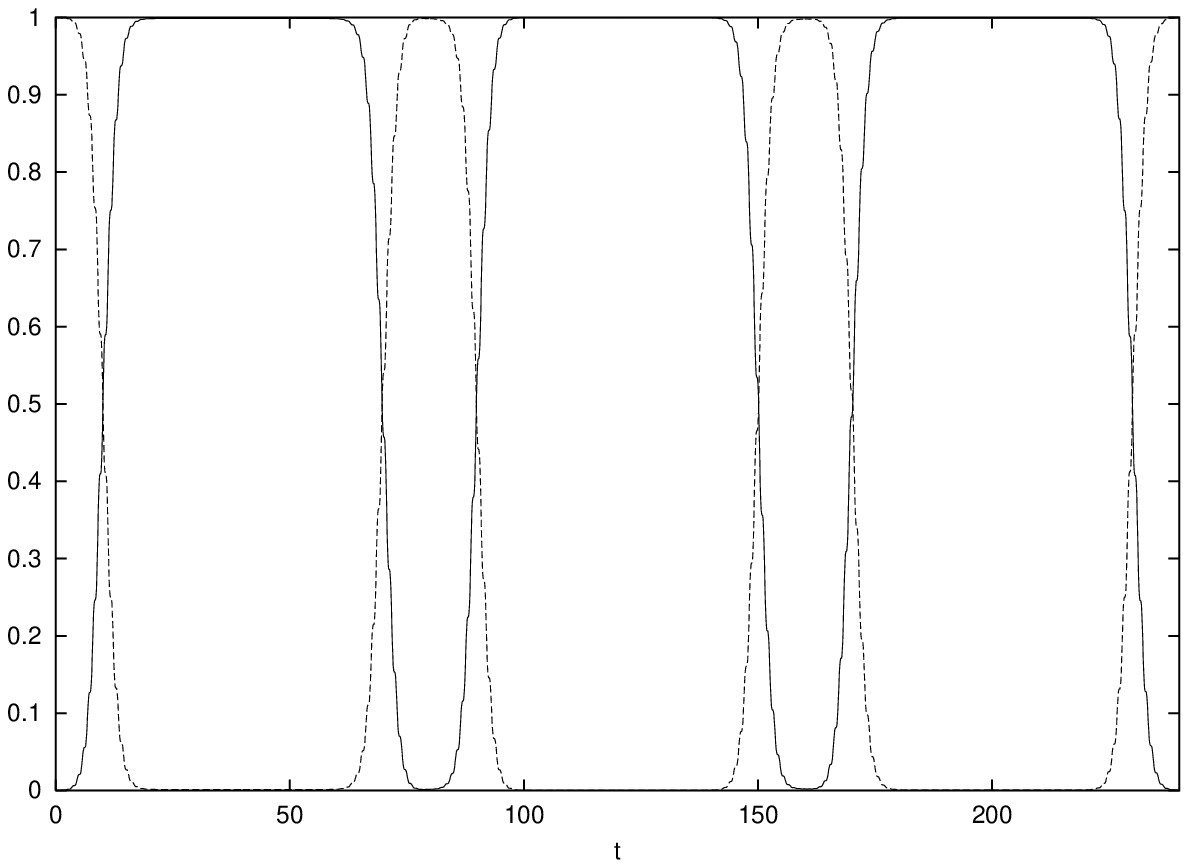,width=3.15in,height=2.5in}}
\subfigure[$\rho_{00}^{(1)}$ and $\rho_{11}^{(1)}$]{
\epsfig{file=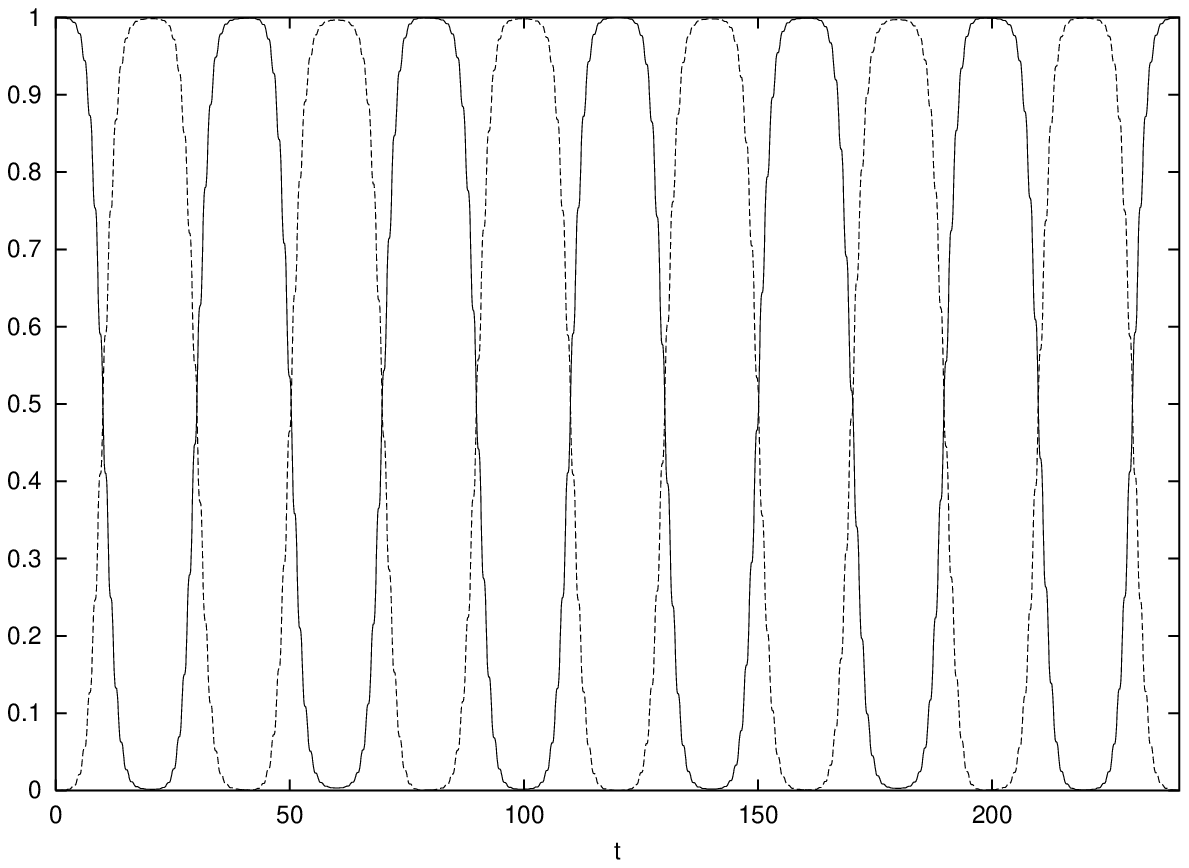,width=3.15in,height=2.5in}}
\subfigure[$\rho_{00}^{(2)}$ and $\rho_{11}^{(2)}$]{
\epsfig{file=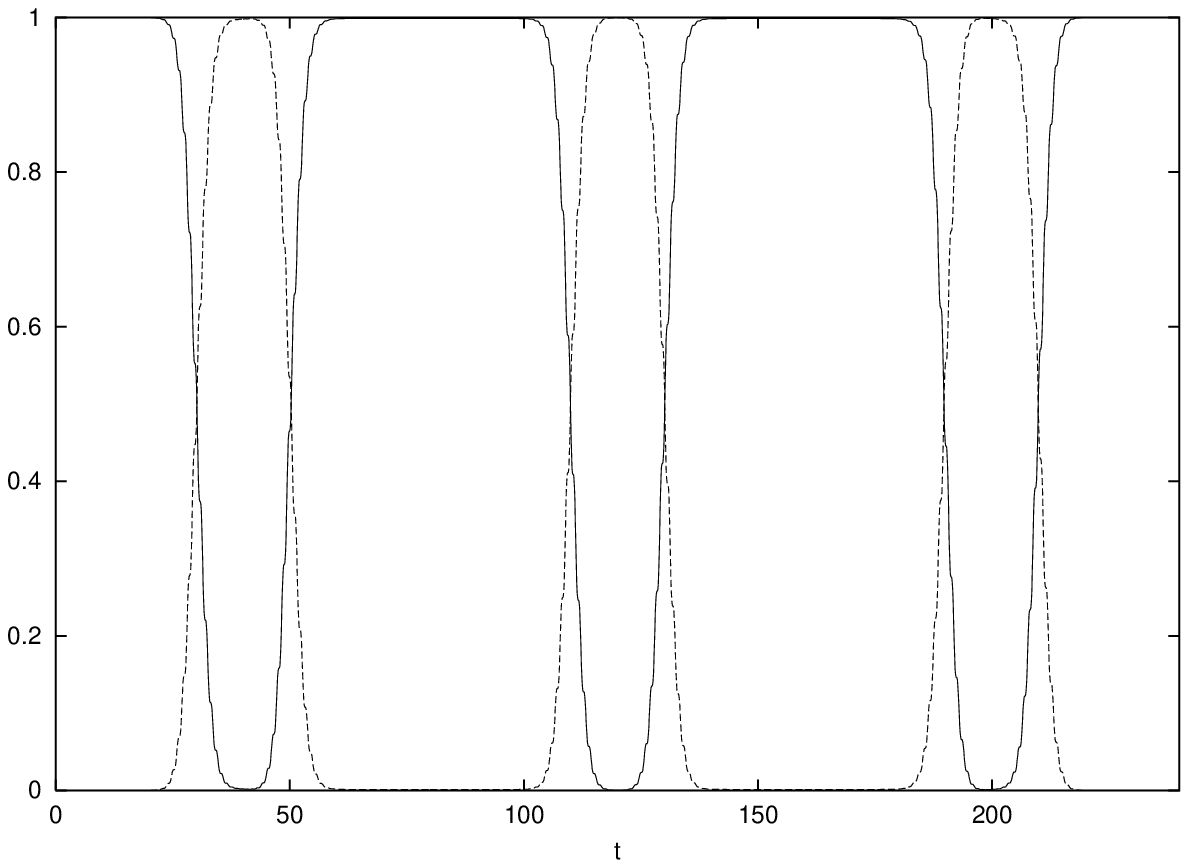,width=3.15in,height=2.5in}}
\end{figure}

We thus employ a time-dependent Hamiltonian of the form
\begin{eqnarray}
H(t)&=&\frac{\hbar\omega_{eg}}{2}\sigma_{z}^{(0)}+\frac{\hbar\omega_p}{2}\sigma_{z}^{(1)} +\frac{\hbar\omega_{eg}}{2}\sigma_{z}^{(2)}+ \hbar\lambda_0(\sigma_{x}^{(0)}+\sigma_{x}^{(1)}+\sigma_{x}^{(2)})\sum_{j=1}^{n_s}\sigma_{x}^{(j+2)}\nonumber \\
&+& \sum_{j=1}^{n_s}[\frac{\hbar\omega_{j+2}}{2}\sigma_{z}^{(j+2)}+\hbar\beta\sigma_{x}^{(j+2)}]+\hbar\lambda\sum_{i=1}^{n_s-1}\sum_{j=i+1}^{n_s}\sigma_{x}^{(i+2)}\sigma_{x}^{(j+2)}\nonumber \\
&+&\hbar a \sum_{k=0}^2\{\sigma_{x}^{(0)}\sigma_{x}^{(1)}e^{-b(t-t_1-k\tau)^2}+\sigma_{x}^{(1)}\sigma_{x}^{(2)}e^{-b(t-t_2-k\tau)^2}\nonumber \\
&+&\sigma_{x}^{(1)}\sigma_{x}^{(2)}e^{-b(t-t_3-k\tau)^2}
+\sigma_{x}^{(0)}\sigma_{x}^{(1)}e^{-b(t-t_4-k\tau)^2}\}\cos \omega_{laser}t
\label{HAM2}
\end{eqnarray}
where we chose the gap between the electronic states of the impurities (labeled 0 and 2) to be $\hbar\omega_{eg}=3$ eV. The excitation energy of the optical phonon (labeled 1) was set to $\hbar\omega_p=.2$ eV. A large system-bath coupling of $\hbar\lambda_0=.0075$ eV was chosen so that decoherence could be observed during the roughly 160 fs of time evolution. An intra-bath coupling of $\hbar\lambda=.03$ eV was chosen which is roughly representative of diamond. A small anharmonic term with $\hbar\beta=.0001$ eV was included. Bath frequencies were sampled from a Debye distribution with a cutoff at $\hbar\omega_c=.05$ eV.

The laser interactions move an initial excitation from the first impurity to the optical phonon and then from the optical 
phonon to the second impurity. The process is then reversed. Overall we repeat this sequence
three times. The parameters of the lasers are $\omega_{laser}=\omega_{eg}-\omega_p$ (less than the diamond band gap of 5.4 eV), $\hbar a=.325$ eV, $b=.325 a^2$. Finally, the first pulse sequence times are
$t_1=10$ $\hbar/$eV, $t_2=30$ $\hbar/$eV, $t_3=50$ $\hbar/$eV and $t_4=70$ $\hbar/$eV with multiples delayed by $\tau=80$ $\hbar/$eV.

We calculated the reduced density of the three qubit system via the formula
\begin{equation}
\rho(t)=\sum_{m=1}^{n_{eig}}p_m ~{\rm Tr}_{b}\{|\psi_m(t)\rangle\langle \psi_m(t)|\}\label{DENSB}
\end{equation}
where
\begin{eqnarray}
p_m=\exp\{-\epsilon_m/kT\}/\sum_{l=1}^{n_{eig}}\exp\{-\epsilon_l/kT\},
\end{eqnarray} 
$\epsilon_m$ and $|m\rangle$ are bath energies and eigenvectors (i.e. of terms 5 and 6 of Eq. (\ref{HAM2})), and $kT=.0067$ eV is the (liquid nitrogen) temperature in units of energy. The notation ${\rm Tr}_{b}\{|\psi_m(t)\rangle\langle \psi_m(t)|\}$ indicates a trace of the full density $|\psi_m(t)\rangle\langle \psi_m(t)|$ over the environmental degrees of freedom. The states $|\psi_m(t)\rangle$ are evolved via the Schr\"{o}dinger equation from initial states
\begin{eqnarray}
|\psi_m(0)\rangle=|100\rangle\otimes |m\rangle\label{istate}
\end{eqnarray}
under Hamiltonian (\ref{HAM2}). The basis of eigenstates of the $\sigma_z$ operators was used to represent all states. The state $|100\rangle$ in (\ref{istate}) refers to the system and means that the 0-spin was initially excited while the 1-spin and 2-spin were in their ground states. 

The ARPACK linear algebra software\cite{Arp} was used to calculate the lowest $n_{eig}=10$ energies and eigenvectors of the isolated environment. The temperature was chosen such that no states with quantum number $m$ higher than $n_{eig}$ are populated at equilibrium. The numerical solutions of the Schr\"{o}dinger ordinary differential equations for $|\psi_m(t)\rangle$ were calculated using an eighth order Runge-Kutta routine\cite{RK}. Operations of the Hamiltonian (\ref{HAM2}) on the statevector were calculated via repeated application of Pauli matrix multiplication routines. For example 
\begin{eqnarray}
\langle j_0,j_1,\dots,j_i,\dots,j_{n_s+2}|\sigma_{x}^{(i)}|\psi\rangle=\langle j_0,j_1,\dots,\bar{j_i},\dots,j_{n_s+2}|\psi\rangle
\end{eqnarray}
for all sets of $j_l=0,1$, $l=0,1,\dots ,n_s+2$ and where $\bar{j_i}=1$ if $j_i=0$ and $\bar{j_i}=0$ if $j_i=1$. Thus, an operation of $\sigma_{x}^{(i)}$ simply rearranges the components of $|\psi\rangle$. States of the basis 
can be represented by integers $j=j_0+j_12+j_22^2+\dots+j_i2^i+\dots+j_{n_s+2}2^{n_s+2}$ and since integers are represented in binary form on a computer, the mapping $j\rightarrow j'=j_0+j_12+j_22^2+\dots+\bar{j_i}2^i+\dots+j_{n_s+2}2^{n_s+2}$ under $\sigma_{x}^{(i)}$ can be calculated very simply
using Fortran binary-operation system functions. Operations for $\sigma_{y}^{(i)}$ and 
$\sigma_{z}^{(i)}$ are also straightforward. 

Finally, from the reduced density (\ref{DENSB}) of the system we calculated the reduced densities of the two qubits and of the optical phonon mode by tracing out the remaining unwanted degrees of freedom. For example, for qubit 0 we calculated
\begin{equation}
\rho^{(0)}(t)={\rm Tr}_1{\rm Tr}_2\{\rho(t)\}
\end{equation}
while for qubit 2 and for phonon mode 1
\begin{eqnarray}
\rho^{(2)}(t)&=&{\rm Tr}_0{\rm Tr}_1\{\rho(t)\}\\
\rho^{(1)}(t)&=&{\rm Tr}_0{\rm Tr}_2\{\rho(t)\}.
\end{eqnarray}
Thus, each component of the system is represented by a 2$\times$2 matrix which makes it easier to display the solutions and compare them with solutions of the master equation.

In order to show convergence to the master equation results we considered a range of values of $n_s$. Specifically, we report results for $n_s=10$, 12, 14 and 16.

For reference we plot the solutions for the subsystem in the absence of dissipation in Fig. 1. Figure 1(a) shows the occupation probabilities for the ground state (solid curve) and excited state (dashed) of the first impurity plotted against time in units of $\hbar/$eV$=.66$ fs. The real and imaginary parts of $\rho_{01}^{(0)}$ are identically zero. Similar quantities for the optical phonon and the second impurity are plotted in 1(b) and 1(c), respectively. Again the off-diagonal elements are zero as a consequence of our choice of initial state.

\section{Numerical solution of master equation}

We recently developed a numerical technique for solving integro-differential equations\cite{TU} like (\ref{masterb}). The accuracy of the method has been established for both generalized Langevin
equations and master equations by comparison with exact solutions\cite{TU}. Basically the method works by
converting integro-differential equations to ordinary differential equations which are solved by standard methods.

\begin{figure}[htp]
\caption{Impurity 1 with dissipation}
\subfigure[$\rho_{00}^{(0)}$]{
\epsfig{file=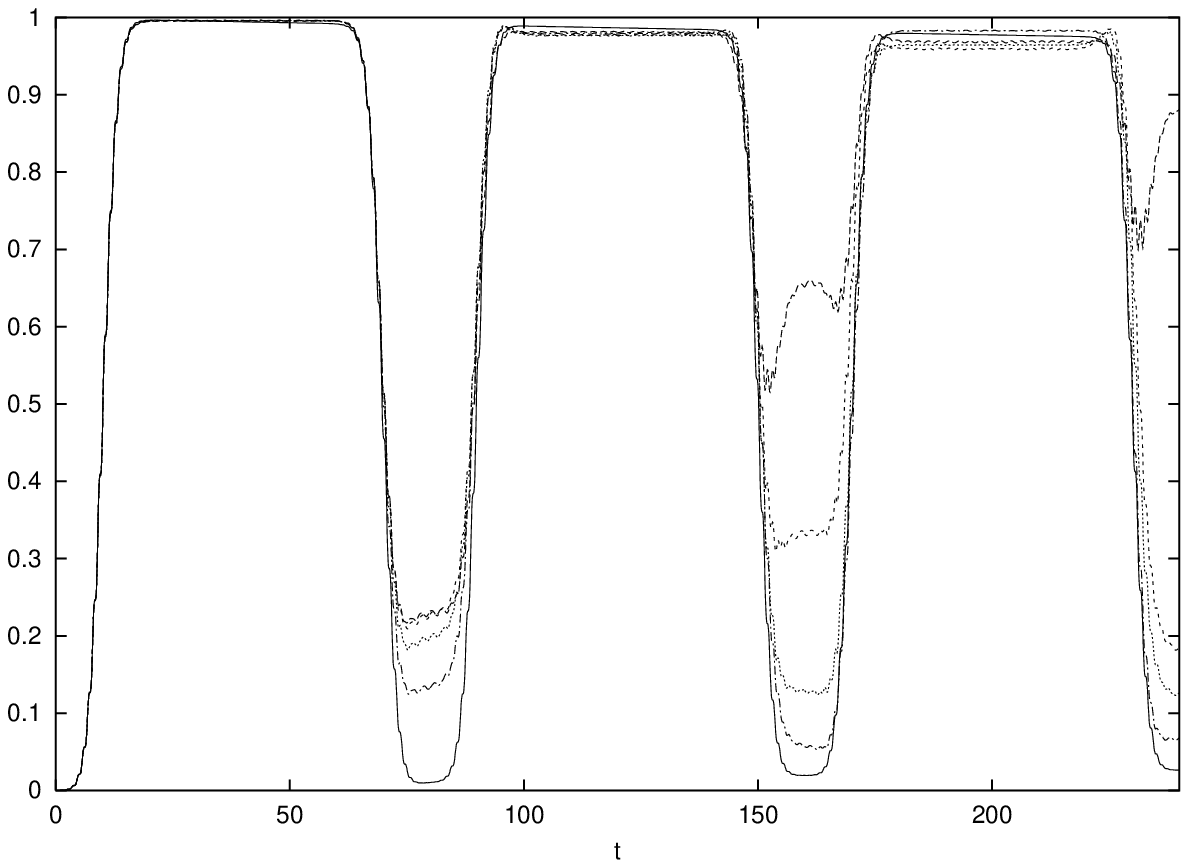,width=3.15in,height=2.5in}}
\subfigure[$\rho_{11}^{(0)}$]{
\epsfig{file=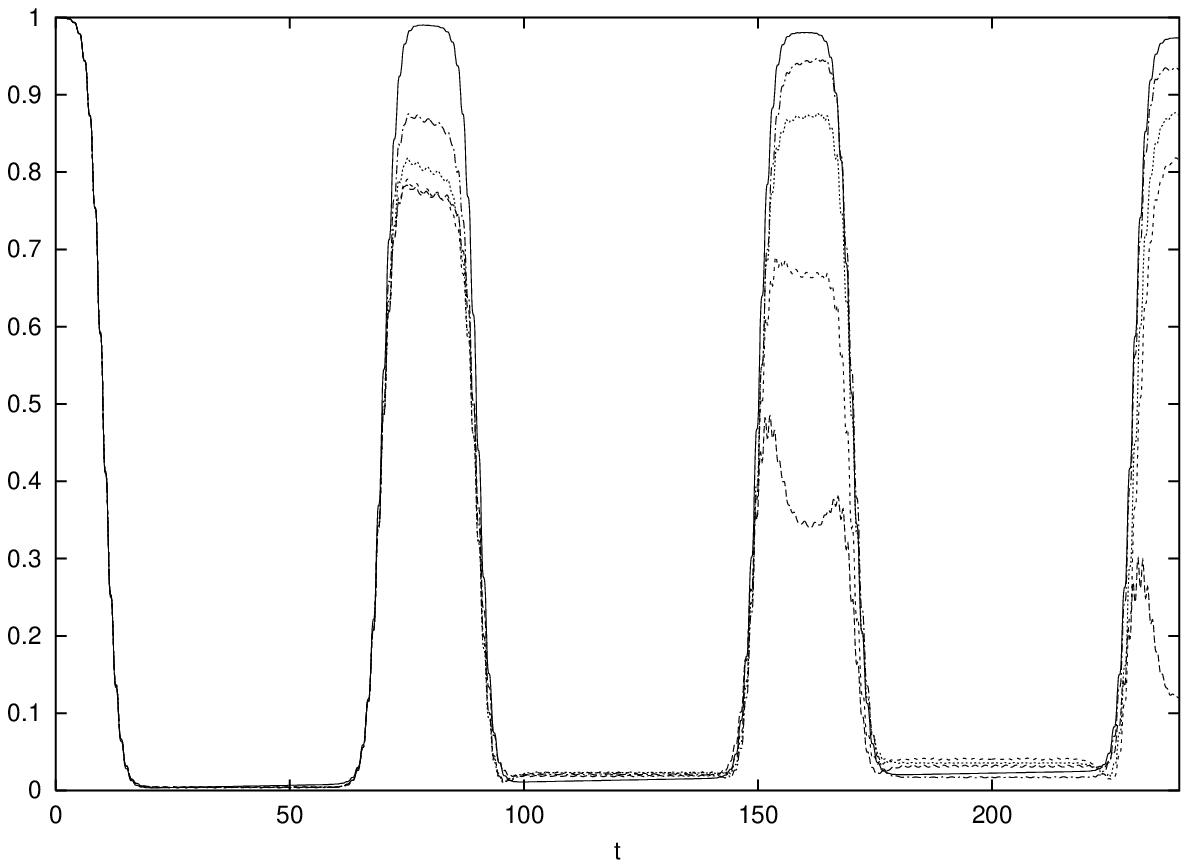,width=3.15in,height=2.5in}}
\subfigure[${\rm Re}\{\rho_{01}^{(0)}\}$]{
\epsfig{file=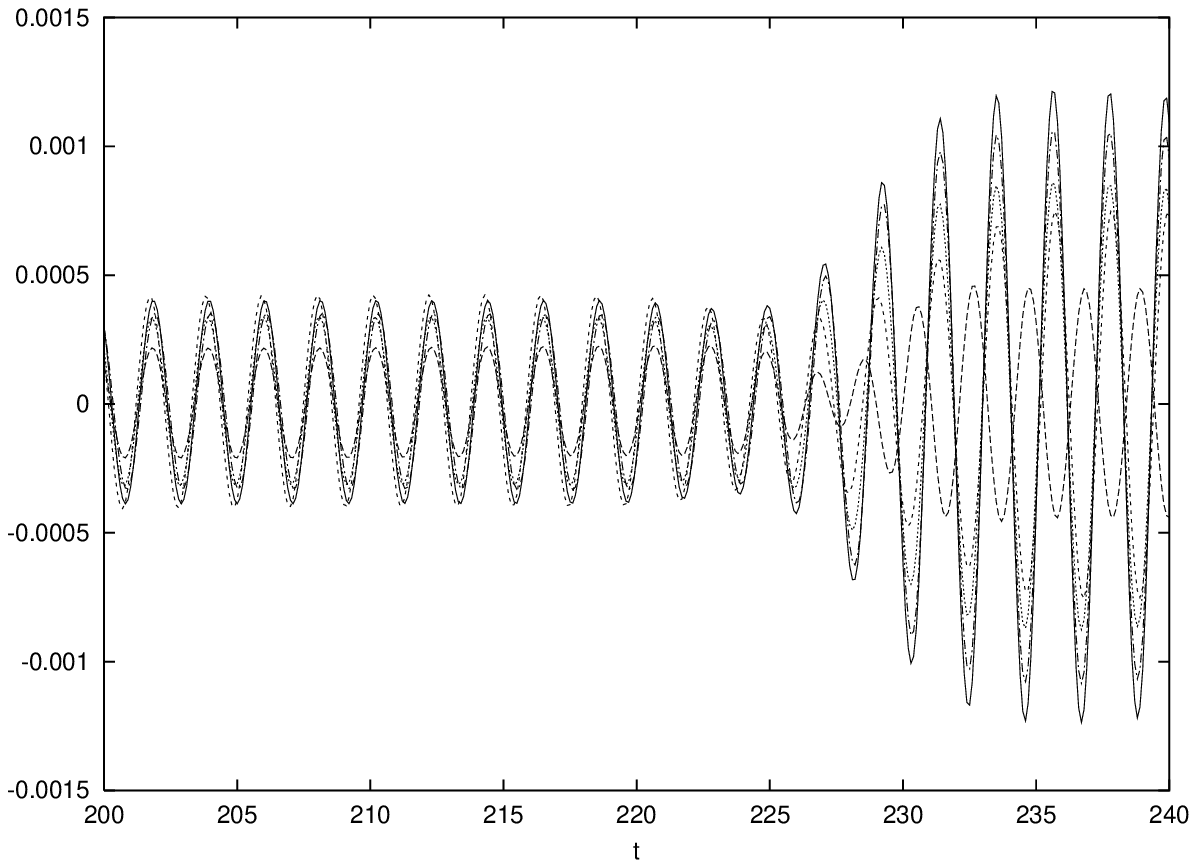,width=3.15in,height=2.5in}}
\subfigure[${\rm Im}\{\rho_{01}^{(0)}\}$]{
\epsfig{file=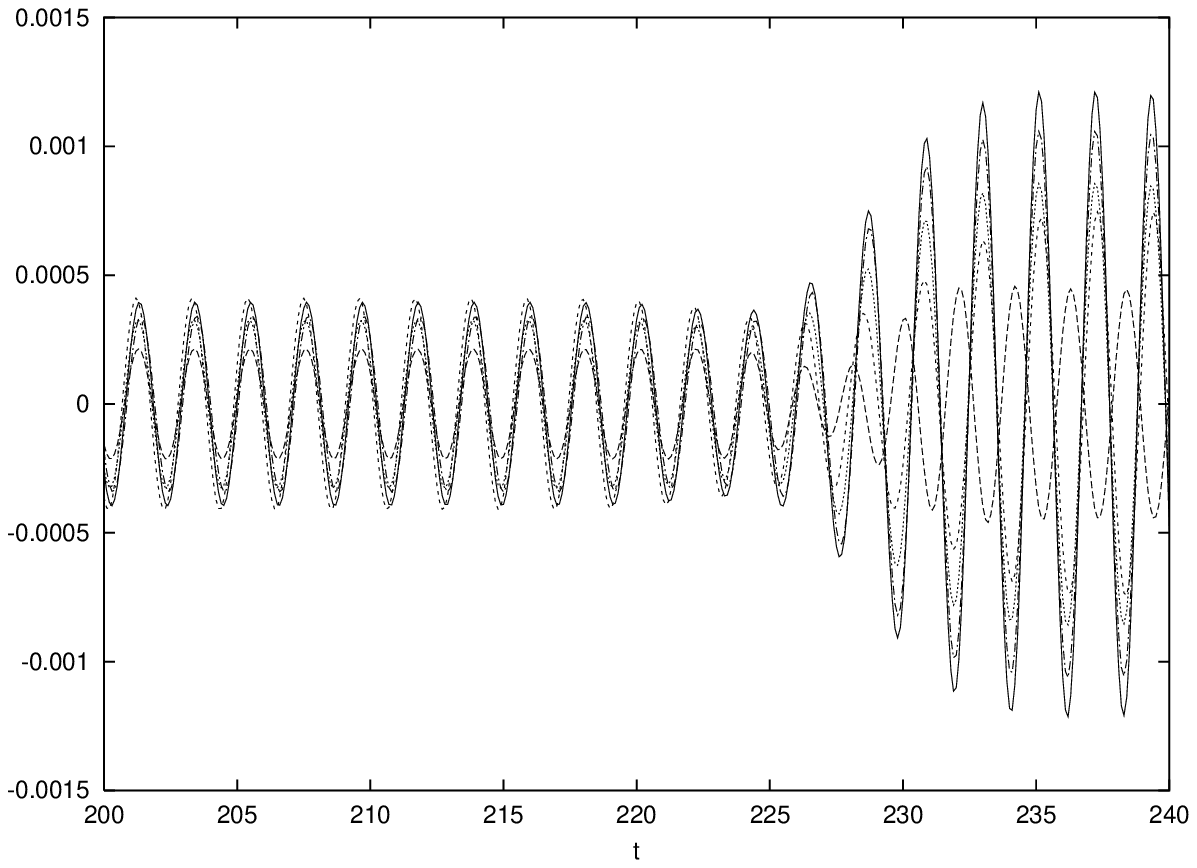,width=3.15in,height=2.5in}}
\end{figure}

We implement the method
as follows. Define a space-like time variable $u$ and a smoothed density operator
\begin{eqnarray}
\chi(t,u)=f(u)\int_0^tdt'~W(t-t'+u)\rho(t'),
\label{SMO}
\end{eqnarray}
where $f(u)$ is a damping function such that $f(0)=1$. Direct substitution shows that $\rho(t)$ and $\chi(t,u)$ satisfy ordinary differential equations
\begin{eqnarray}
&&d\rho(t)/dt=-(i/\hbar)[H_s+{\cal E}(t)+\sum_{\mu}\bar{R}_{\mu}S_{\mu},\rho(t)]\nonumber \\
&&~~~~~~~~~~~~-(1/\hbar^2)\sum_{\mu,\nu}C_{\mu,\nu}\{[\chi(t,0)S_{\nu},S_{\mu}]+
[S_{\nu},S_{\mu}\chi(t,0)]\},\label{tu1}\\
&&d\chi(t,u)/dt=f(u)W(u)\rho(t)+\partial \chi(t,u)/\partial u-(f'(u)/f(u))~\chi(t,u)\label{tu2}
\end{eqnarray}
(where $f'(u)=df(u)/du$) which are then solved by representing $u$ on a grid.

More specifically, the equations for Hamiltonian (\ref{HAM2}) are
\begin{eqnarray}
d\rho(t)/dt&=&-i[\omega_{eg}/2\sigma_{z}^{(0)}+\omega_p/2\sigma_{z}^{(1)} +\omega_{eg}/2\sigma_{z}^{(2)}+ \lambda_0(\sigma_{x}^{(0)}+\sigma_{x}^{(1)}+\sigma_{x}^{(2)})\bar{\Sigma}_x \nonumber \\
&+&a \sum_{k=0}^2\{\sigma_{x}^{(0)}\sigma_{x}^{(1)}e^{-b(t-t_1-k\tau)^2}+\sigma_{x}^{(1)}\sigma_{x}^{(2)}e^{-b(t-t_2-k\tau)^2}\nonumber \\
&+&\sigma_{x}^{(1)}\sigma_{x}^{(2)}e^{-b(t-t_3-k\tau)^2}
+\sigma_{x}^{(0)}\sigma_{x}^{(1)}e^{-b(t-t_4-k\tau)^2}\}\cos \omega_{laser}t
,~\rho(t)]\nonumber \\
&-&C\{(\sigma_{x}^{(0)}+\sigma_{x}^{(1)}+\sigma_{x}^{(2)})^2\chi(t,0)+\chi(t,0)(\sigma_{x}^{(0)}+\sigma_{x}^{(1)}+\sigma_{x}^{(2)})^2\nonumber \\
&-&2(\sigma_{x}^{(0)}+\sigma_{x}^{(1)}+\sigma_{x}^{(2)})\chi(t,0)(\sigma_{x}^{(0)}+\sigma_{x}^{(1)}+\sigma_{x}^{(2)})\}\label{stu1}\\
d\chi(t,u)/dt&=&e^{-g u^2}W(u)\rho(t)+\partial \chi(t,u)/\partial u+2gu~\chi(t,u)\label{stu2}
\end{eqnarray}
where $\bar{\Sigma}_x={\rm Tr}_b\{\Sigma_x{\cal B}\}$, $\Sigma_x=\sum_{j=1}^{n_s}\sigma_{x}^{(j+2)}$, and $C=\lambda_0^2{\rm Tr}_b\{\Sigma_x^2{\cal B}\}$.

Equations (\ref{stu1}) and (\ref{stu2}) were restricted to a grid of points $u_j=(n+l-j)\Delta t$ with $j=1,\dots, n$ and $l=int(.338n)$ where $\Delta t=.1$ $\hbar/$eV is the time-step employed in the dynamics. Following Ref. \cite{TU} a damping function $f(u)=e^{-g u^2}$ with $g=  11/[(n-l)\Delta t]^2$ was used. Converged results were obtained for $n=100$ grid points. We chose $W(u)=W(|u|)$ for negative values of $u$. A discrete-variable\cite{DVR} matrix representation 
was employed to calculate the partial derivative with respect to $u$ in Eq. (\ref{stu2}). 
\begin{figure}[htp]
\caption{Optical phonon with dissipation}
\subfigure[$\rho_{00}^{(1)}$]{
\epsfig{file=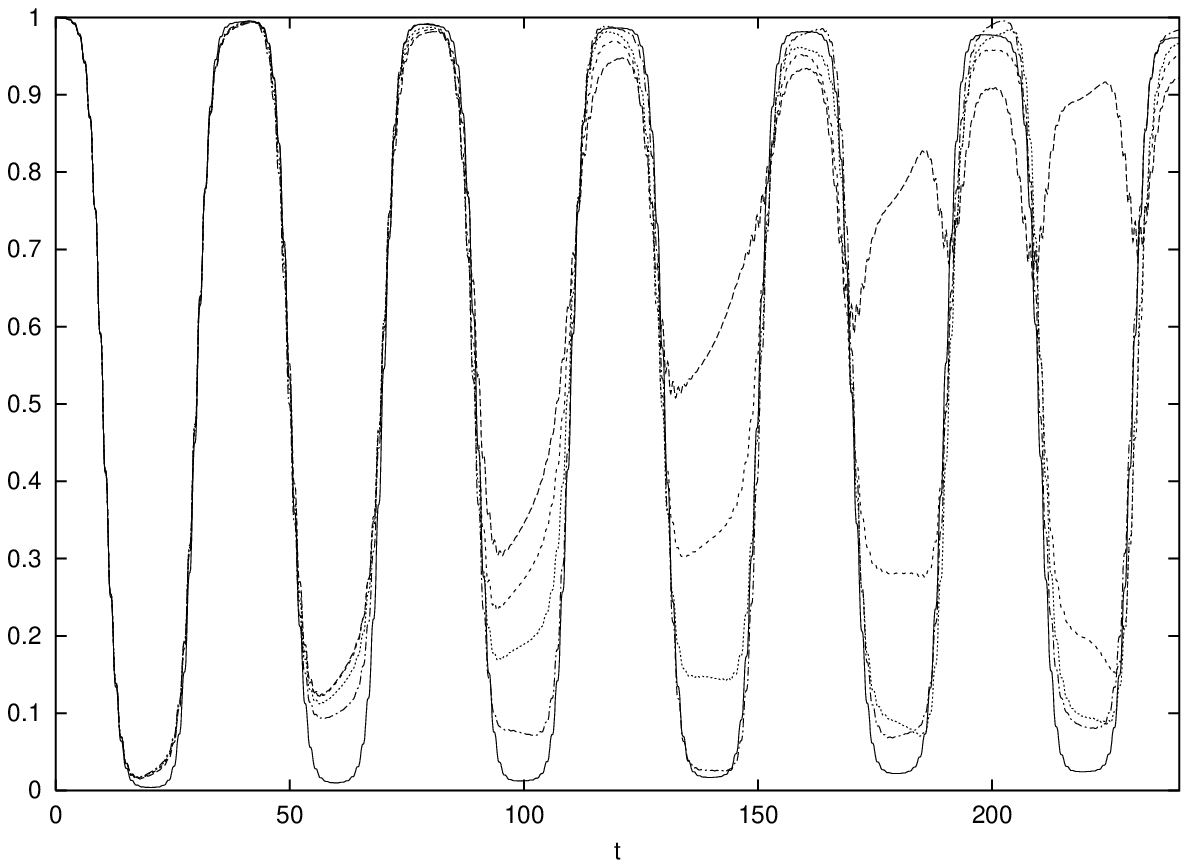,width=3.15in,height=2.5in}}
\subfigure[$\rho_{11}^{(1)}$]{
\epsfig{file=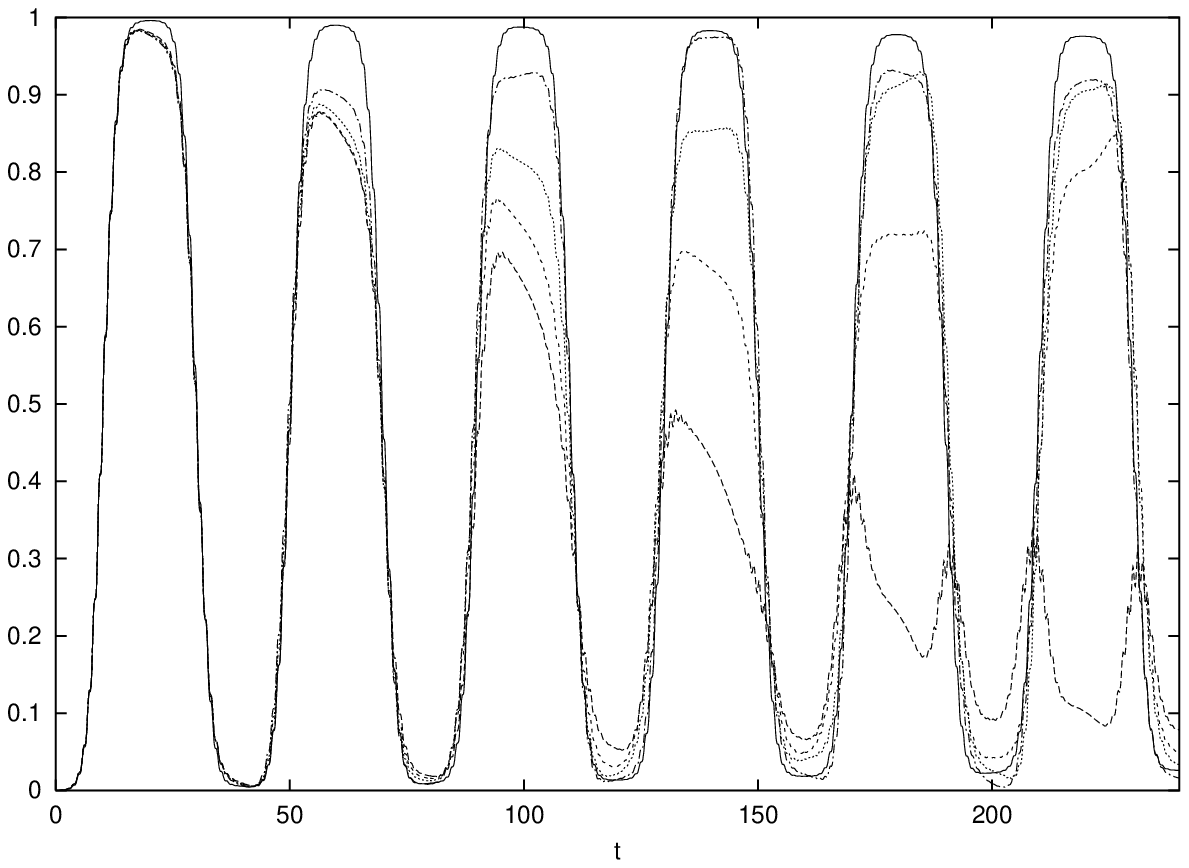,width=3.15in,height=2.5in}}
\subfigure[${\rm Re}\{\rho_{01}^{(1)}\}$]{
\epsfig{file=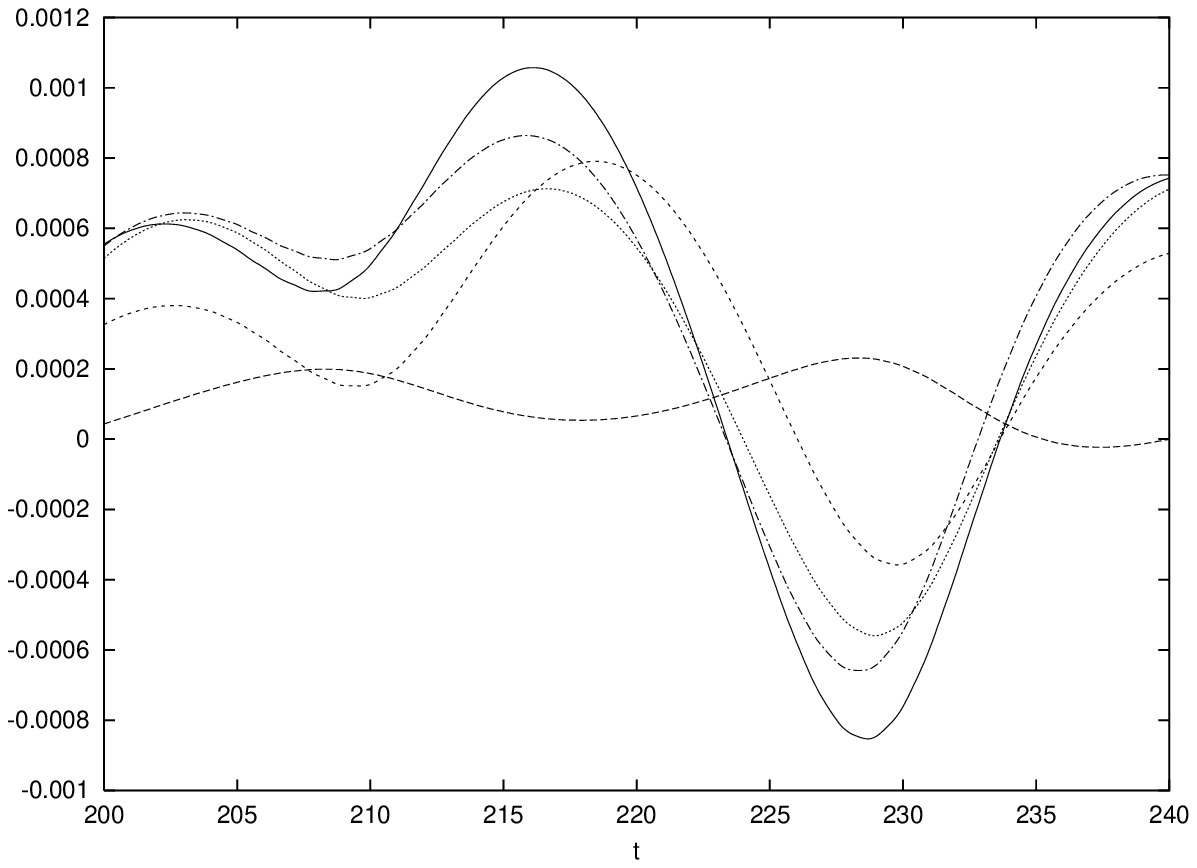,width=3.15in,height=2.5in}}
\subfigure[${\rm Im}\{\rho_{01}^{(1)}\}$]{
\epsfig{file=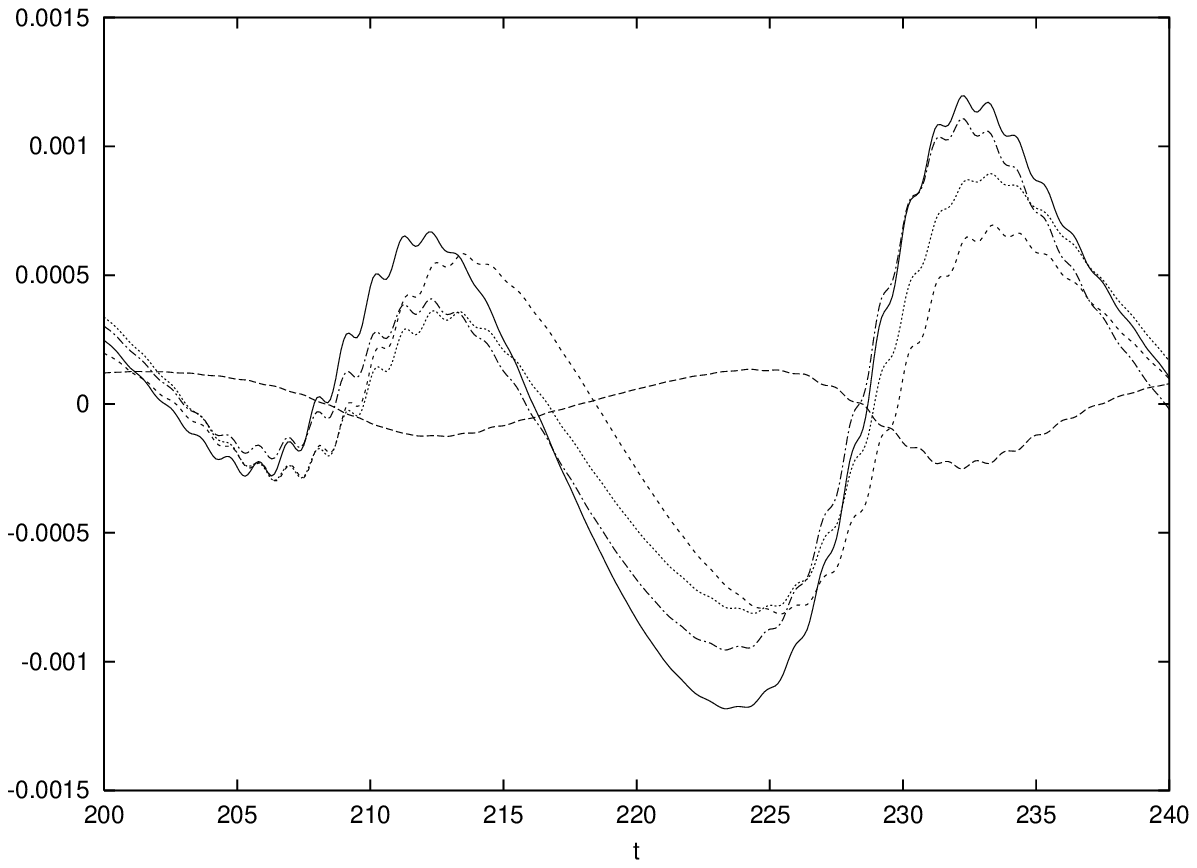,width=3.15in,height=2.5in}}
\end{figure}

Finally, the ordinary differential equations (\ref{stu1}) and (\ref{stu2}) were integrated using an eighth order Runge-Kutta routine\cite{RK}. 

The parameters of the master equation $\bar{\Sigma}_x$ and $C$, and the parameters of the memory function (\ref{W2}), were calculated using the exact energies and eigenvectors of the bath Hamiltonian computed in Section III and formulas reported in Ref. \cite{JJW}. The parameters of the master equation converge rapidly with the number of bath spins since they are all average quantities. Thus, master equation solutions for $n_s=12$ cannot be distinguished from solutions with $n_s=14$ or $n_s=16$. 

\section{Results}

Figures 2-4 show the occupation probabilities and real and imaginary parts of off-diagonal density elements for each of the three qubits, for various values of the number of bath modes $n_s$, as a function of time. Curves for $n_s=10$ (dashed), $n_s=12$ (short-dashed), $n_s=14$ (dotted), and $n_s=16$ (dot-dashed) are shown in each figure. For comparison we also show the solutions of the master equation (solid curve). [Note that the occupation probabilities for the master equation are positive, in agreement with our proof in section III.] Time $t$ is in units of $\hbar/$eV$\simeq .66$ fs.

Each sequence of four laser pulses can be viewed as moving an excitation from the first impurity to the optical phonon and from the phonon to the second impurity, then back to the phonon and finally back to the first impurity. The sequence is repeated three times for a total of twelve laser pulses. The general idea
is to roughly simulate the sort of manipulations that would be employed in a quantum computer. Because calculation of the system dynamics in the presence of the bath spins is very expensive we have chosen a strong system-bath coupling and short pulse width so that decoherence is manifested over the relatively short time span of 160 fs.

In accord with the initial conditions and pulse sequence Fig. 2 shows an excitation on impurity 1 ($\rho^{(0)}_{11}=1$, $\rho^{(0)}_{00}=0$), which relaxes to its ground state ($\rho^{(0)}_{11}=0$, $\rho^{(0)}_{00}=1$) and then is re-excited. This is repeated three times. Figure 3 shows the phonon mode initially in its ground state ($\rho^{(1)}_{11}=0$, $\rho^{(1)}_{00}=1$). The phonon is then excited ($\rho^{(1)}_{11}=1$, $\rho^{(1)}_{00}=0$) and then returned to its ground state. Again, this is repeated three times. The second impurity, shown in Fig. 4, is initially in its ground state ($\rho^{(2)}_{11}=0$, $\rho^{(2)}_{00}=1$). It is then excited ($\rho^{(2)}_{11}=1$, $\rho^{(2)}_{00}=0$) and then returned to its ground state. The real and imaginary parts of the off-diagonal elements of the three qubits show small oscillations throughout the manipulations as a result of decoherence. These oscillations are much faster for the two impurities and so we show only the last forty time units. Compare Fig. 1 with Figs. 2-4 and note the obvious effects of decoherence and dissipation present in the dynamics of all density components. Return to the the initial state is imperfect because of decoherence. Note also that while the off-diagonal elements are zero in the absence of dissipation, here they show small oscillations.
\begin{figure}[htp]
\caption{Impurity 2 with dissipation}
\subfigure[$\rho_{00}^{(2)}$]{
\epsfig{file=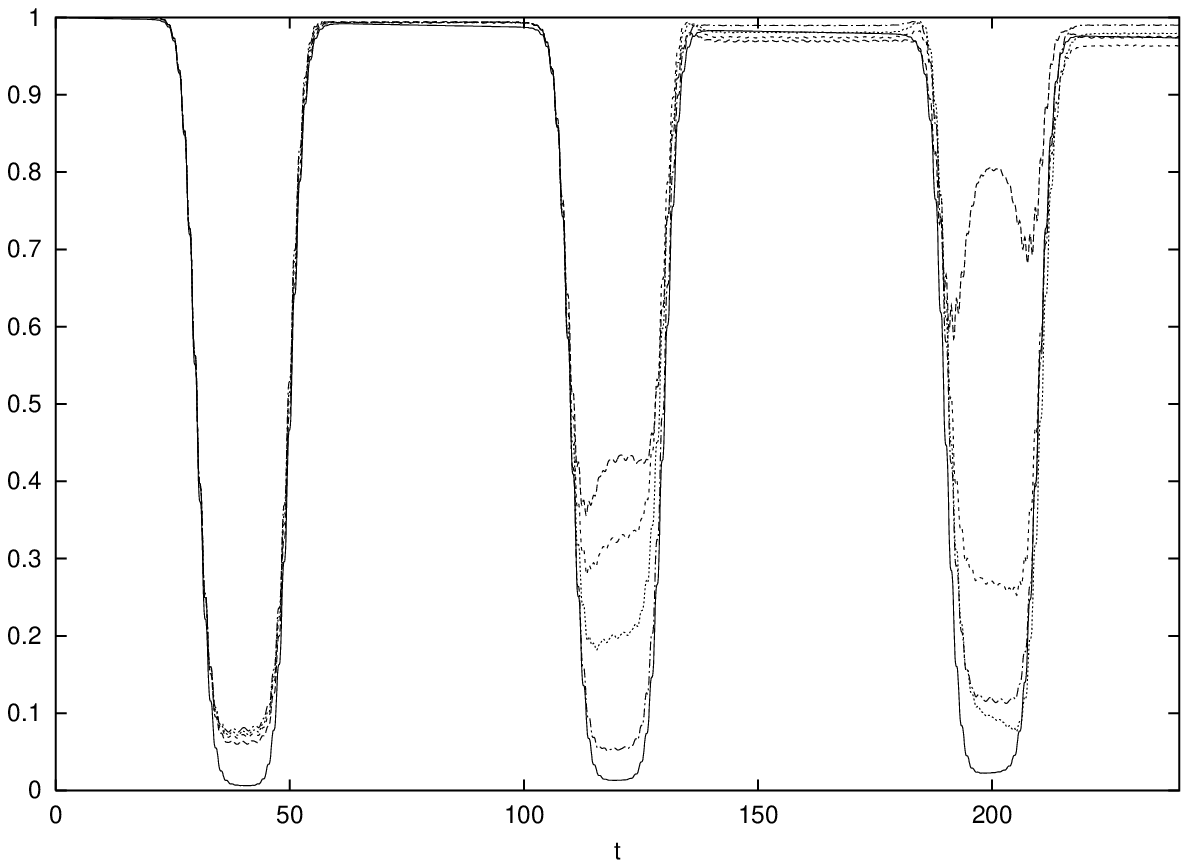,width=3.15in,height=2.5in}}
\subfigure[$\rho_{11}^{(2)}$]{
\epsfig{file=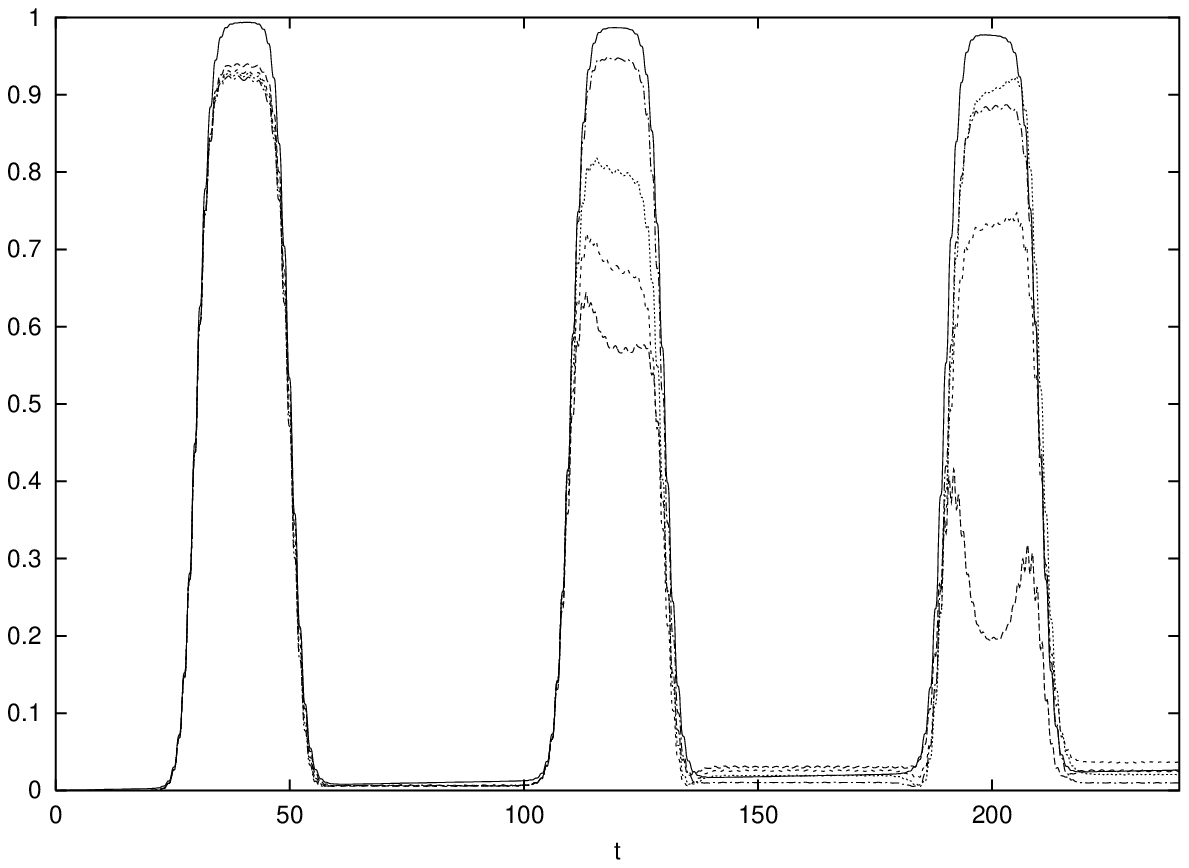,width=3.15in,height=2.5in}}
\subfigure[${\rm Re}\{\rho_{01}^{(2)}\}$]{
\epsfig{file=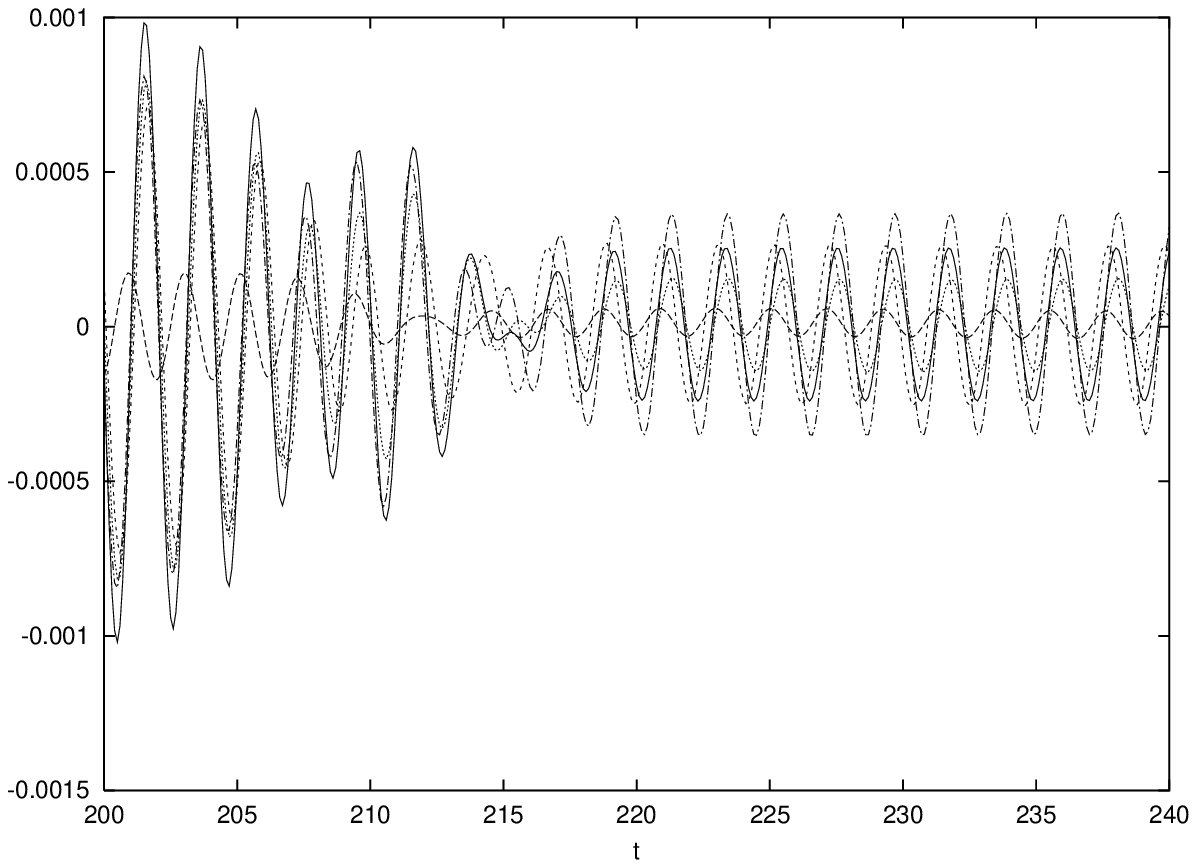,width=3.15in,height=2.5in}}
\subfigure[${\rm Im}\{\rho_{01}^{(2)}\}$]{
\epsfig{file=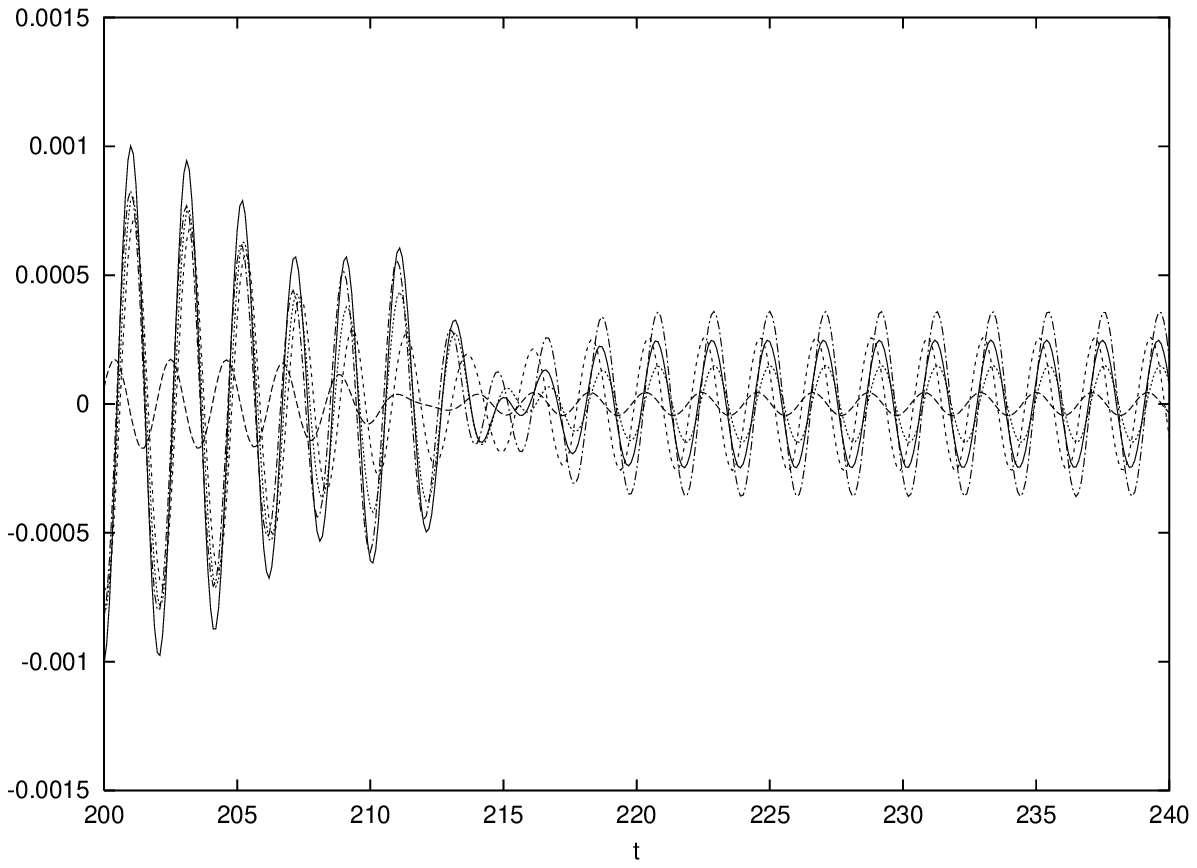,width=3.15in,height=2.5in}}
\end{figure}

It is clear that the agreement between the exact calculations and the master equation improves dramatically with increasing numbers of bath spins. Decoherence effects are generally much stronger for small baths but these decrease as the bath gets larger. For small numbers of bath spins, the return to the initial state after a pulse sequence is less perfect than that predicted by the master equation. As the number of bath spins increases this discrepancy is incrementally reduced. Deviations of the real and imaginary parts of the off-diagonal elements from master equation predictions also decline as the number of bath spins increases. 

The results for 16 bath spins show close but still imperfect agreement with the predictions of the master equation. It is obviously of interest whether the exact numerical solutions converge {\em precisely} to the master equation results in the limit that the bath is very large. Unfortunately, we were unable to perform calculations for larger baths and so this will have to remain an open question.

\section{Summary}

We have used a mean field approximation to derive a master equation suitable for time-dependent subsystem Hamiltonians and self-interacting baths. After proving that the master equation preserves positivity of the reduced density matrix 
we compared its solutions to those of a model system. We found that exact numerical solutions for the model converged toward those of the master equation as the number of bath modes was increased. This supports our expectation that the 
approximate master equation will become increasing accurate as the bath size approaches the thermodynamic limit.

We have recently developed an exact method for decomposing the quantum N-body vibronic dynamics 
problem (for pairwise interaction) into N stochastic 1-body problems\cite{PRE}. That is, we can now exactly solve the dynamics of pairwise interacting distinguishable spins and vibrations. This should allow us to obtain exact solutions for more realistic models and for larger baths. We hope to soon test the predictions of the mean field master equation against exact solutions for these more realistic models.

The authors gratefully acknowledge the support of the Natural Sciences and 
Engineering Research Council of Canada.


\begin{thebibliography}{99}

\bibitem{OLD} See D. Kohen, C.C. Marston and D.J. Tannor, J. Chem. Phys. 107, 5236 (1997); P. Gaspard and M. Nagaoka, J. Chem. Phys. 111, 5668 (1999); U. Weiss, {\em Quantum dissipative systems, 2nd Ed.}, (World Scientific, Singapore, 1999).

\bibitem{Gasp} P. Gaspard, M.E. Briggs, M.K. Francis, J.V. Sengers, R.W. Gammon, J.R. Dorfmann and R.V. Calabrese, Nature 394, 865 (1998).

\bibitem{Miyano} T. Miyano, S. Munetoh, K. Moriguchi and A. Shintani, Phys. Rev. E 64, 016202 (2001).

\bibitem{TW} L. Tessieri and J. Wilkie, quant-ph/0209079, submitted for publication.

\bibitem{JJW} J. Wilkie, quant-ph/0306087, submitted for publication.

\bibitem{JW} J. Wilkie, Phys. Rev. E 62, 8808 (2000).

\bibitem{Wilk1} J. Wilkie, J. Chem. Phys. 114, 7736 (2001).

\bibitem{Wilk2} J. Wilkie, J. Chem. Phys. 115, 10335 (2001).

\bibitem{Zwan} S. Nakajima, Prog. Theor. Phys. 20, 948 (1958); R. Zwanzig,
J. Chem. Phys. 33, 1338 (1960); R. Zwanzig, in {\em Lectures in Theoretical Physics}, 
Vol. 3 (Interscience, New York, 1961).

\bibitem{Opp} P. Gaspard and M. Nagaoka, J. Chem. Phys. 111, 5668 (1999); A. Su\'{a}rez, R. Silbey and I. Oppenheim,
  J. Chem. Phys. 97, 5101 (1992); V. Romero-Rochin and I. Oppenheim,
  J. Stat. Phys. 53, 307 (1988); Physica A 155, 52 (1989);
  V. Romero-Rochin, A. Orsky and I. Oppenheim, {\em ibid.} 156, 244 (1989).

\bibitem{WB} See J. Wilkie and P. Brumer, Phys. Rev. A 61, 064101 (2000), and references therein for a discussion of Dirac notation for Liouville-Hilbert space.

\bibitem{Pesk} U. Peskin and N. Moiseyev, J. Chem. Phys. 99, 4590 (1993); P. Pfeifer and R.D. Levine, J. Chem. Phys. 79, 5512 (1983).

\bibitem{dsg} G. Lindblad, Commun. Math. Phys. 48, 119 (1976); V. Gorini, A. Kossakowski, and E.C.G. Sudarshan, J. Math. Phys. 
17, 821 (1976); R. Alicki and K. Lendi, {\em Quantum Dynamical Semigroups and
Applications}, (Springer, Berlin, 1987).

\bibitem{Arp} See http://www.caam.rice.edu/software/ARPACK/.

\bibitem{RK} DOP853.f, E. Hairer and G. Wanner,\\ http://elib.zib.de/pub/elib/hairer-wanner/nonstiff/.

\bibitem{DVR} D.T. Colbert and W.H. Miller, J. Chem. Phys. 96, 1982 (1992).

\bibitem{TU} J. Wilkie, quant-ph/0306088, accepted for publication in Phys. Rev. E, 2003.

\bibitem{PRE} J. Wilkie, Phys. Rev. E 67, 017102 (2003).

\end{thebibliography}
\end{document}